# EFFECT OF INSECURITY ON AGRICULTURAL OUTPUT IN BENUE STATE, NIGERIA

[1]Victor Ushahemba Ijirshar, [2]Isaiah Iortyom Udaah, [3]Bridget Ngodoo Mile, [4]Joyce Seember Vershima, & [5]Abba Adaudu

[1,2,3,4]Department of Economics, Benue State University, Makurdi-Nigeria
[2]Department of Economics, Federal University of Lafia-Nigeria

**Abstract**

This study examined the effect of insecurity on agricultural output in Benue state. A descriptive survey design was employed, and 400 respondents were purposively selected from insecurity-prone local government areas, namely, Guma LGA, Agatu LGA, Gwer LGA, Gwer-West LGA, Katsina-Ala LGA, Logo LGA, Ukum LGA and Kwande LGA. The data were collected through the administration of a questionnaire and were analysed using t tests and structural equation modelling (SEM). The t-test was used to compare farmers' incomes before and after the insecurity in the study area to assess if the differences were statistically significant, while Structural Equation Modelling analysed the complex relationships among multiple variables, employing regression and factor analysis to model both direct and indirect effects. The results revealed that the monetary value of crop and livestock output decreased during periods of insecurity. Furthermore, the study showed that insecurity has an adverse effect on crop and livestock output. This means that a one percent increase in insecurity leads to a 0.211% and 0.311% decrease in crop and livestock output respectively. The study concluded that insecurity reduced agricultural output in Benue state. Based on the findings, it was recommended that the government deploy more security personnel, establish community policing initiatives, and employ modern surveillance technologies to deter criminal activities in insecure areas. Additionally, for places experiencing farmer-herder conflict, the government should provide grazing reserves for herdsmen and further strengthen the state law on open grazing prohibition and the establishment of ranch law.

# INTRODUCTION

Agriculture is the backbone of many economies around the world, particularly in developing countries where it serves as the primary source of livelihood, employment, and sustenance for a significant portion of the population. It plays a critical role in ensuring food security, reducing poverty, and fostering economic growth (Usman, 2022; Abubakar, 2021). The economy of Benue State is largely based on agriculture, which has been the primary occupation of its people since ancient times. Initially, early inhabitants of the state relied on the natural resources of the land for sustenance. However, as population pressure increased and these natural resources became scarce, humans began domesticating animals and cultivating plants. Consequently, agriculture is defined as the science of cultivating crops and rearing animals for human use

(Ijirshar, Ker, & Terlumun, 2015). However, the agricultural sector is increasingly facing several challenges that threaten its productivity and sustainability. Among these challenges, insecurity has emerged as a paramount concern with far-reaching implications.

Insecurity, encompassing a spectrum of issues from armed conflicts and banditry to theft, vandalism, and communal clashes, disrupts agricultural activities, leading to decreased productivity, loss of income, and heightened food insecurity. Ewetan and Urhie (2014) defined insecurity as a breach of peace and security, whether historical, religious, ethno-regional, civil, social, economic, or political, that contributes to recurring conflicts and leads to wanton destruction of lives and property. For more than a decade, the Nigerian state has been confronted with the monstrous malaise of a security crisis. The pattern and trend of insecurity in Nigeria in recent years have ranged from insurgency, banditry, kidnapping activities, and communal clashes to herder-farmer conflicts. These issues are particularly pronounced in regions that are predominantly agricultural, such as the Middle Belt, which includes Benue State, and the northern parts of the country (Omitola, Adedire, Akinrinde, Omodunbi & Sackflame, 2021). The causes of insecurity are complicated and involve competition over natural resources, ethnic tensions, political instability, and economic disparities.

These threats disrupt agricultural activities, leading to reduced crop yields, livestock losses, and abandoned farmlands. The Food and Agriculture Organization (FAO) (2021) has highlighted that insecurity is a leading cause of food insecurity, as it disrupts food production, supply chains, and market access. This effect has been confirmed empirically by Musa, Salami and Umoru (2022), Nomor and Ikyoyer (2021), Abubakar (2021), Eneji, Babagario and Agri (2019), and Ijirshar, Ker, and Terlumun (2015). However, the Nigerian government, alongside the Benue state government, has implemented various measures to address insecurity and bolster agricultural productivity. These efforts include the deployment of security task forces such as Operation Whirl Stroke, Operation Ayem a Kpatuma (a military operation aimed at tackling the menace), livestock guard, local vigilante groups to protect rural areas, and the enactment of the Benue State open grazing prohibition and ranches establishment law (Benue State Government, 2017). Despite these efforts, insecurity still persists in Benue State. As such, this study examines the effect of insecurity on agricultural output in Benue State.

# LITERATURE REVIEW

## Theoretical framework

The study is based on conflict theory and the Cobb–Douglas production function. Conflict theory, originally developed by Karl Marx, posits that society is in a state of perpetual conflict due to competition for limited resources. This theory has evolved over time to encompass various dimensions of social life, including economic disparities, political power struggles, and social inequalities. Insecurity, whether in the form of political instability, social unrest, or economic uncertainty, can create barriers to agricultural production. In societies where there is a lack of security, farmers may face challenges such as land grabs, theft of crops or livestock, destruction of infrastructure, or displacement due to conflict. These disruptions can lead to decreased agricultural output and food shortages, exacerbating poverty and inequality within communities.

The Cobb–Douglas production function is a widely used economic model that describes the relationship between inputs and outputs in a production process. It takes the form $Q = AL^{\alpha}K^{\beta}$, where Q represents the output, L stands for labour input, K represents capital input, A is total factor productivity, and $\alpha$ and $\beta$ are the output elasticities of labour and capital, respectively. The primary ways in which insecurity impacts agricultural output through the lens of the Cobb–Douglas production function include first, insecure environments often leading to the displacement of populations, reducing available labour for agricultural activities. Farmers may abandon their fields due to safety concerns, leading to a decrease in labour. Reduced labour input directly lowers agricultural output, and insecurity can result in the destruction or loss of capital assets such as machinery, irrigation systems, storage facilities, and livestock. This reduction in capital negatively affects agricultural productivity. Additionally, insecurity can deter investment in new capital due to heightened risk.

## Empirical Review

Saad (2024) investigated the effects of insecurity on agricultural production in the Batsari Local Government Area of Katsina State, revealing a positive correlation between insecurity and agricultural production through multiple regression analysis. Similarly, Usman (2022) explored the impact of insecurity on food production in the Igabi Local Government Area of Kaduna State, finding that kidnapping, banditry, and cattle rustling contribute to food shortages and rising prices. These studies highlight the detrimental effects of insecurity on local agricultural economies, stressing the role of insecurity in disrupting food production in affected regions. In a similar vein, Nwosu, Ndukwe, Aguwamba, and Uchegbu (2023) investigated the impact of human-induced global warming and insecurity on agricultural productivity in Nigeria from 1981 to 2020 using the ARDL method. Their findings revealed that population growth and methane emissions are positively correlated with agricultural productivity, whereas insecurity, characterized by terrorist attacks and fatalities, negatively affects agricultural output.

Further expanding on this, Abubakar (2021) examined Nigeria's agricultural output from 1999 to 2020, showing that inflation negatively affects agricultural performance, while government spending on internal security, unemployment, corruption perception, and trade openness positively influence it. Similarly, Eneji, Babagario, and Agri (2019), using OLS regression in Balanga LGA, Gombe State, found an inverse relationship between crime, unemployment, and agricultural productivity. Complementing these findings, Ijirshar, Ker, and Terlumun (2015) focused on the farmer-herder conflicts in Benue State, revealing that frequent attacks by Fulani Herdsmen severely reduced agricultural output and had broader socioeconomic consequences, including reduced crop yields, income, displacement, and widespread destruction of infrastructure. These studies collectively underscore the multifaceted impact of insecurity on Nigeria's agricultural sector.

## METHODOLOGY

A descriptive survey design was employed for the study. The design involved gathering data and systematically describing the typical features or facts about a certain community from a few people or items thought to be representative of the entire group. Eight (8) local governments were identified as being adversely affected by insecurity in Benue State for inclusion in the study. These include the Guma LGA, Agatu LGA, Gwer LGA, Gwer-West LGA, Katsina-Ala LGA, Logo LGA, Ukum LGA and Kwande LGA. The selection and inclusion of these local governments for the study of the impact of insecurity on agricultural output are well justified due to the prevalence of various types of insecurity in these areas. Each of these local government areas (LGAs) has experienced significant security challenges that have directly affected agricultural activities, which is a crucial aspect of the study. For instance, Guma LGA (farmer-herder conflicts, banditry) has been one of the epicentres of farmer-herder conflicts, which have resulted in loss of lives, displacement of farmers, and destruction of farmlands; Agatu LGA (ethnic clashes, farmer-herder conflicts) has witnessed intense clashes between local communities and herders, leading to significant agricultural setbacks; Gwer LGA (communal clashes, farmer-herder conflicts) has experienced communal clashes and conflicts between farmers and herders, affecting agricultural activities; similar to Gwer, Gwer-West LGA (farmer-herder conflicts, communal violence) has faced frequent farmer-herder conflicts and communal violence; Katsina-Ala LGA (banditry, kidnapping, communal clashes) is known for its high incidence of banditry and kidnappings, which have created an environment of fear and instability; Logo LGA (armed banditry, farmer-herder conflicts) has been plagued by armed banditry and conflicts between farmers and herders; and the Ukum LGA (banditry, communal conflicts) has experienced banditry and communal conflicts, disrupting the agricultural sector. This insecurity has affected farming operations, leading to a decrease in agricultural output and food security concerns.

These selected LGAs represent areas in Benue State that have been significantly affected by various forms of insecurity, including farmer-herder conflicts, banditry, kidnappings, and communal violence. These security challenges have had direct and adverse effects on agricultural output, making these LGAs pertinent for the study of the impact of insecurity on agriculture. The population of the study comprises crops and livestock farmers. A sample size of 400 respondents was determined using the Taro Yamane formula, and a buffer margin of 10% was employed to increase the sample size to 440; 55 respondents were selected from eight local government areas, namely, the Guma LGA, Agatu LGA, Gwer LGA, Gwer-West LGA, Katsina-Ala LGA, Logo LGA, Ukum LGA and Kwande LGA, due to the frequent occurrence of insecurity. The purposive sampling technique was used to select respondents from each local government. Primary data were used for the study and collected with the use of a structured questionnaire.

This study employs a two-pronged statistical approach to rigorously assess the impact of insecurity on agricultural output in Benue State. First, paired-samples t-test was conducted to compare the average annual monetary value of crop output before and during periods of insecurity. This test facilitates a straightforward evaluation of whether there is a significant difference in crop and livestock production attributable to periods of heightened insecurity. Prior to conducting the t-test, assumptions such as normality and homogeneity of variances are thoroughly examined to ensure the validity of the results.

Complementing the t-test, Structural Equation Modeling (SEM) was utilized to assess the relationships between insecurity and both crop and livestock outputs. The SEM framework incorporates a range of factors influencing agricultural productivity, including the number of attacks, number of deaths, level of fear, access to agricultural credit, availability of fertilizer, pest prevalence, farm size, and climate conditions. By modeling both direct and indirect effects, SEM allows for a comprehensive analysis of how these variables interact and collectively impact agricultural output. The model's fit is evaluated using standard indices such as the Normed Fit Index (NFI), Incremental Fit Index (IFI), Comparative Fit Index (CFI), Relative Fit Index (RFI), Tucker–Lewis Index (TLI) and Root Mean Square Error of Approximation (RMSEA) to ensure its robustness. This analytical approach aligns with the methodology used by Abah, Ochoche, and Stephen (2022) in their study on the impact of herder-farmer conflict on rice production.

Equation 1: Effect of insecurity on crop production in the Benue state.

$$CRP = f\left(INS_i, ACAC, AVF, FSIZ, PP\right) \qquad (1)$$

where CRP represents crop production (measured by the monetary value of crop outputs) and INS refers to insecurity (insecurity is a latent variable encompassing various dimensions of insecurity experienced within a community or locality). The latent variable "Insecurity" is measured by six indicators, each assessed using Likert scale responses ranging from "Strongly Agree" (SA) to "Strongly Disagree" (SD) to provide insight into different aspects of insecurity:

INS1: Massive killing.

INS2: Massive destruction of homes and farmlands.

INS3: Frequent attacks in the locality.

INS4: The number of fatalities and injuries resulting from clashes has been increasing.

INS5: Several people were forced to flee their homes due to the conflict.

INS6: Damage to agricultural lands and livestock caused by conflicts.

Together, these indicators provide an assessment of the nature of insecurity within the community, encompassing both human and material dimensions of vulnerability and disruption caused by conflict and violence.

ACAC stands for Access to agricultural credit (measured by the number of times farmers have access to agricultural credit), AVF is the availability of fertilizers (measured by the number of times farmers have access to fertilizers), FSIZ is farm size (measured by the number of hectares of cultivated land) and PP is the prevalence of pests.

Equation 2: Effect of insecurity on livestock production in the Benue state.

$$LVP = f(INS_i, ACAC, AVF, FSIZ, PP) \tag{2}$$

where LVP is livestock production (measured by the monetary value of livestock outputs).

These expressions can be presented in the structural equation model (SEM) as:

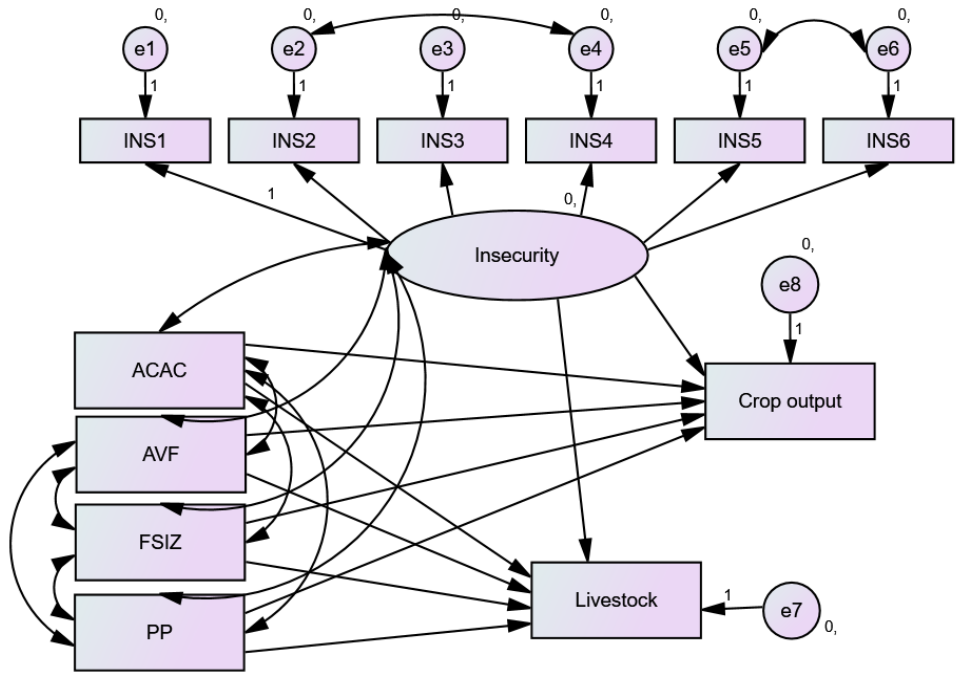

Figure 1: SEM Model for Estimating the Effect of Insecurity on Agricultural Output.

In the confirmatory analysis of the latent variable Insecurity (INS), additional covariance relationships were introduced to improve the model's fit and enhance its explanatory power. Specifically, covariance was allowed between the error terms e2 and e4, as well as between e5 and e6. These covariances reflect correlations between the measurement errors of key indicators and were included to account for potential unobserved factors influencing responses, thereby refining the estimation of the latent variable. The covariance between e2 and e4 captures the relationship between the errors of INS2 (massive destruction of homes and farmlands) and INS4 (increasing fatalities and injuries from clashes). This suggests that factors not directly measured by the model may simultaneously affect responses to these indicators. By incorporating this covariance, the model acknowledges shared variance between the two indicators, which leads to a more precise estimation of the Insecurity construct and improves the overall model fit.

Similarly, the covariance between e5 and e6 represents the relationship between the errors of INS5 (people forced to flee their homes due to conflict) and INS6 (damage to agricultural lands and livestock). Again, this covariance allows the model to account for potential shared sources of variation affecting both indicators, beyond what is captured by the structural equation model (SEM). Including this relationship helps to reduce measurement error and increases the robustness of the model in reflecting the underlying concept of Insecurity. By incorporating these covariance relationships—between e2 and e4, and e5 and e6—the model becomes more accurate in capturing shared variance among indicators, thus improving the precision of the latent variable estimation and refining the overall model fit.

## RESULTS AND DISCUSSION

This section discusses the relationship between insecurity and agricultural output in Benue State. However, agricultural output was disaggregated into crop and livestock farmers in the most affected area with insecurity. The results are shown in Table 1.

**Table 1: Respondents' annual monetary value of crop and livestock output before and during insecurity**

| Annual Monetary Value | Crops Outputs | | | | Livestock Outputs | | | |
| --- | --- | --- | --- | --- | --- | --- | --- | --- |
| | Before Insecurity | | During Insecurity | | Before Insecurity | | During Insecurity | |
| | F | P | F | P | F | P | F | P |
| Below ₦50,000 | 23 | 5.8 | 146 | 37.0 | 18 | 4.5 | 283 | 70.8 |
| ₦50,001 – ₦100,000 | 175 | 43.8 | 190 | 47.5 | 34 | 8.5 | 78 | 19.5 |
| ₦100,001 – ₦500,000 | 184 | 46.0 | 42 | 10.5 | 255 | 63.8 | 26 | 6.5 |
| ₦500,001 Above | 18 | 4.4 | 20 | 5.0 | 93 | 23.2 | 13 | 3.2 |
| Total | 400 | 100 | 400 | 100 | 400 | 100 | 400 | 100 |
| Means | ₦200,770.50 | | ₦124,298.50 | | ₦371,354.75 | | ₦82,219.75 | |
| Means Difference | ₦76,472.000 | | | | ₦289,135.000 | | | |
| t test statistic (Prob.) | ₦19.387(0.000) | | | | ₦30.78(0.000) | | | |

**Source:** Extractions from SPSS 27

Table 1 shows that 5.8% of the respondents reported an annual monetary value below ₦50,000, 43.8% of the respondents reported an annual monetary value between ₦50,001 and ₦100,000, 46.0% of the respondents reported an annual monetary value between ₦100,001 and ₦500,000, 4.4% of the respondents reported an annual monetary value above ₦500,001 before experiencing insecurity in the study area, 37.0% of the respondents reported an annual monetary value below ₦50,000, 47.5% of the respondents reported an annual monetary value between ₦50,001 and ₦100,000, 10.5% of the respondents reported an annual monetary value between ₦100,001 and ₦500,000, and 5.0% of the respondents reported an annual monetary value above ₦500,001. The paired t test results showed that the average annual monetary values of crop outputs before and during insecurity were ₦200,770.50 and ₦124,298.50, respectively, with a mean difference of ₦76,472. The results also revealed a t test statistical value of 19.387 and a probability value of 0.0000<0.05. The null hypothesis that the mean difference between the average annual monetary value of crop output before and during periods of insecurity is zero is rejected at the 5% level of significance.

Furthermore, the monetary value of livestock output results in Table 1 show that 4.5% of the respondents reported an annual monetary value below ₦50,000, 8.5% of the respondents reported an annual monetary value between ₦50,001 and ₦100,000, 63.8% of the respondents reported an annual monetary value between ₦100,001 and ₦500,000, 23.2% of the respondents reported an annual monetary value above ₦500,001 before experiencing insecurity in the study area, 70.8% of the respondents reported an annual monetary value below ₦50,000, 19.5% of the respondents reported an annual monetary value between ₦50,001 and ₦100,000, 6.5% of the respondents reported an annual monetary value between ₦100,001 and ₦500,000, and 3.2% of the respondents reported an annual monetary value above ₦500,001. The paired t test results showed that the average annual monetary values of crop outputs before and during insecurity were ₦371,354.75 and ₦82,219.75, respectively, with a mean difference of ₦289,135. The results also revealed a t-test statistic value of 30.780 and a probability value of 0.0000<0.05. The null hypothesis that the mean difference between the average annual monetary value of crop output before and during periods of insecurity is zero is rejected at the 5% level of significance.

To further evaluate this relationship, structural equation modelling (SEM) was employed, and the results are shown in Table 2.

**Table 2: SEM Results on the Impact of Insecurity on Agricultural Output in Benue State**

|  | Unstandardized Estimates | Standardized Estimates | S.E. | C.R. | P | Lower | Upper | P |
|---|---|---|---|---|---|---|---|---|
| INS1<---Insecurity | 1 | 0.984 |  |  |  | 0.974 | 0.993 | 0.002 |
| INS2<---Insecurity | 0.707 | 0.776 | 0.03 | 23.865 | *** | 0.719 | 0.825 | 0.001 |
| INS3<---Insecurity | 1.063 | 0.987 | 0.013 | 81.174 | *** | 0.977 | 0.994 | 0.002 |
| INS4<---Insecurity | 0.85 | 0.906 | 0.021 | 39.674 | *** | 0.874 | 0.935 | 0.001 |
| INS5<---Insecurity | 1.153 | 0.969 | 0.018 | 63.816 | *** | 0.957 | 0.982 | 0.001 |
| INS6<---Insecurity | 1.108 | 0.97 | 0.017 | 64.23 | *** | 0.952 | 0.983 | 0.002 |
| CRP<---ACAC | 0.233 | 0.296 | 0.07 | 3.326 | *** | 0.136 | 0.461 | 0.002 |
| CRP<---AVF | 0.39 | 0.6 | 0.077 | 5.101 | *** | 0.359 | 0.829 | 0.001 |
| CRP<---FSIZ | 0.099 | 0.234 | 0.016 | 6.223 | *** | 0.173 | 0.297 | 0.001 |
| CRP<---PP | -0.093 | -0.149 | 0.064 | -1.448 | 0.148 | -0.353 | 0.08 | 0.189 |
| LVP<---ACAC | 0.573 | 0.775 | 0.042 | 13.619 | *** | 0.652 | 0.902 | 0.001 |
| LVP<---AVF | 0.191 | 0.313 | 0.046 | 4.151 | *** | 0.187 | 0.439 | 0.002 |
| LVP<---FSIZ | 0.013 | 0.032 | 0.01 | 1.324 | 0.186 | -0.014 | 0.076 | 0.196 |
| LVP<---PP | 0.01 | 0.017 | 0.039 | 0.261 | 0.794 | -0.113 | 0.159 | 0.751 |
| LVP<---Insecurity | -0.394 | -0.311 | 0.048 | -8.196 | *** | -0.395 | -0.247 | 0.000 |
| CRP<---Insecurity | -0.285 | -0.211 | 0.08 | -3.566 | *** | -0.35 | -0.096 | 0.001 |

**Model Fit Indices: CMIN=604.842 (P=0.000), NFI=0.926, RFI=0.871, IFI=0.93, TLI=0.879, CFI=0.93, RMSEA=0.063, Standardized RMR = .0330**

Source: Extracts from SPSS Amos

The estimated standardized coefficient of insecurity on crop output (CRP) is -0.211, which is statistically significant at the 5% level (P=0.001). This means that a one percent increase in insecurity leads to a 0.211% decrease in crop output in Benue State. The negative relationship is driven by farmer displacement, disrupted farming activities, and reduced access to inputs and markets, which lower crop production, contributing to food insecurity, reduced incomes, and increased poverty among farmers. This finding conforms to the findings of Usman (2022), Musa, Salami and Umoru (2022), Abubakar (2021), Eneji, Babagario and Agri (2019), and Ijirshar, Ker, and Terlumun (2015) but is inconsistent with the findings of Saad (2024).

The estimated standardized coefficient of insecurity on livestock output (LVP) is -0.311, which is statistically significant at the 5% level (P=0.000). This coefficient indicates a negative and significant impact of insecurity on livestock farming in Benue State, implying that a one percent increase in insecurity results in a 0.311% decrease in livestock output. The negative relationship can be explained through various mechanisms, including the displacement of livestock farmers, loss of livestock due to theft or violence, and disruption of essential veterinary and support services. This finding is in line with the findings of Musa, Salami and Umoru (2022), Nomor and Ikyoyer (2021), Abubakar (2021), Eneji, Babagario and Agri (2019), and Ijirshar, Ker, and Terlumun (2015) but is at odds with the findings of Saad (2024).

The estimated standardized coefficients of access to agricultural credit (ACAC), availability of fertilizer (AVF) and farm size (FSIZ) had positive effects on crop production, while the prevalence of pests had a negative effect on crop production. However, only the effect of farm size on crop production is significant at the 5% critical level. On the other hand, the estimated standardized coefficients of access to agricultural credit (ACAC), availability of fertilizer (AVF), farm size (FSIZ), and prevalence of pests are positively related to livestock production (LVP) and statistically significant at the 5% critical level.

The goodness of fit of the structural equation model (SEM) was evaluated using various decision rule thresholds. The chi-square (CMIN) value was significant ($p < 0.05$), indicating a significant difference between the observed and model-implied covariance matrices. Although a nonsignificant chi-square is desirable, it is rare in practice, especially for large sample sizes such as this study. The normed fit index (NFI) was 0.926, suggesting a good fit of the model to the data. The incremental fit index (IFI) was 0.93, indicating a good fit. The comparative fit index (CFI) was 0.93, indicating a good fit of the model to the data. The relative fit index (RFI) and Tucker–Lewis index (TLI) were 0.871 and 0.879, respectively, which are below the acceptable thresholds, suggesting that the model fit could be improved. The root mean square error of approximation (RMSEA) of 0.063 is slightly above the threshold for a good fit but still within the range for a reasonable fit, suggesting that the model approximates the data relatively well. Again, the standardized root mean square residual (SRMR) of 0.0330 suggested an excellent fit of the model to the data.

**CONCLUSION AND RECOMMENDATIONS**

The study reveals that insecurity, characterized by violence and conflict, severely impacts agriculture by displacing farmers from their lands and disrupting their ability to cultivate crops and care for livestock. This displacement reduces agricultural productivity and creates a cycle of fear and uncertainty, causing farmers to postpone critical farming activities and potentially abandon agriculture altogether. Additionally, insecure environments hinder access to essential resources for farming and limit farmers' ability to reach markets, further diminishing their income and economic stability. Overall, these factors collectively lead to a decline in both crop and livestock production, underscoring the detrimental effects of insecurity on agricultural systems and rural economies. This finding implies that insecurity negatively affects agricultural output. Based on this finding, it was recommended that:

i. The government, particularly at both the state and federal levels, should prioritize the enhancement of security in rural areas that are vital for agricultural activities by not only deploying a greater number of security personnel to these regions to ensure a robust law enforcement presence, but also by establishing community policing initiatives that actively engage local residents in safeguarding their communities; furthermore, the utilization of modern surveillance technologies, including drones and closed-circuit television (CCTV) systems, should be integrated into the security framework to monitor agricultural zones effectively and deter potential criminal activities that threaten farming operations.

ii. To combat the pervasive issue of insecurity in the State, the government, in collaboration with law enforcement agencies, must significantly increase the number of security personnel allocated to the most affected areas to ensure that law enforcement is both visible and effective; this initiative should be accompanied by intensified efforts to arrest, prosecute, and impose appropriate penalties on individuals responsible for violent crimes, thereby establishing a specialized task force focused on rural security, which could further bolster these efforts by addressing specific threats faced by agricultural communities.

iii. In regions that are experiencing persistent conflicts between farmers and herders, it is imperative for the State government to take decisive action by creating designated grazing reserves; in addition, the government should reinforce existing laws that prohibit open grazing and actively support the establishment of ranches, thereby contributing to a more structured and peaceful coexistence between farmers and herders.